\documentclass[aps,preprint,onecolumn,showpacs,preprintnumbers,amsmath,amssymb,prb,sort&compress]{revtex4}

\usepackage{graphicx}

\usepackage{bm}

\begin{document}

\title{Suppression of Magnetic Frustration by Doping in a nearest-neighbour anti-ferromagnetic triangular lattice}

\author{K. Sheshadri$^{1}$}
\email{kshesh@gmail.com}
\author{A. Chainani$^{2}$}
\email{chainani@spring8.or.jp}

\affiliation{$^{1}$Bagalur, Bangalore North Taluk, Karnataka - 562149, India}
\affiliation{$^{2}$RIKEN/Spring8, Mikazuki-cho, Hyogo 679-5148, Japan}

\date{\today}

\begin{abstract}
Based on experimental observations in A$_{x}$MO$_{2}$ (A = Na, Li; M = Co, Ni), a model for suppression of magnetic frustration by electron doping in a nearest-neighbour antiferromagnetic triangular lattice is presented. It is found that frustration can be quantified, as determined by geometry and bond-counting, and its magnitude is a non-monotonic function of $x$. A mean-field calculation provides temperature-dependent magnetization, spin-entropy and heat capacity. Low-doping ($x = 0.25, ~0.33$) results in a highly frustrated regime.  A$_{0.5}$MO$_{2}$ has strongest order and no frustration, while high doping ($x = 0.67, ~0.75$) leads to low frustration and higher spin-entropy. The results agree with experiments including neutron scattering, spin entropy-driven thermoelectricity and ion ordering. 

\end{abstract}


\maketitle

The discovery of superconductivity in Na$_{x}$CoO$_{2}.y$ H$_{2}$O has reignited the importance of electrical and magnetic properties of the doped triangular lattice \cite{Takada,Schaak}. 
The structure consists of a layered  triangular lattice of Co ions within oxygen octahedra, with Na ions and water molecules intercalated between these layers \cite{Takada,Schaak}.
The Na$_{x}$CoO$_{2}$ ($x \sim$ 0.6 ~ - ~ 0.8) compounds are important materials for thermoelectric applications as they exhibit a large thermopower, low resistivity and low thermal conductivity \cite{Terasaki,Wang}. Systematic experiments \cite{Chou1,Huang1,Amatucci,Delmas,Zandbergen} and theory \cite{Arroyo,Pickett,Zhang,Roger} as a function of $x$ have identified charge, spin, and alkali metal-ion (A) ordered phases at particular $x$ values in the isostructural family of compounds A$_{x}$MO$_{2}$ (A = Na, Li; M = Co, Ni). These include the material used in present-day rechargeable Lithium batteries \cite{Tarascon}. Strong correlations (on-site Coulomb energy $\sim$3-5 eV in the cobaltates \cite{Pickett,vanElp,Chainani}) are important for spin-charge order, thermodynamics and transport properties. Doping, frustration and fluctuations 
in a triangular lattice
(TL) lead to well-known results, including the resonating valence bond state and superconductivity \cite{Anderson, Ogata, Maekawa}.

Na$_{0.5}$CoO$_{2}$ undergoes a charge and magnetic ordering
metal-insulator transition with $T$ \cite{Huang1,Yokoi,Gasparovic}. For $x < 0.5$ a paramagnetic correlated metal phase exists, with a significantly enhanced in-plane  magnetic susceptibility $\chi$ \cite{Chou1}. In contrast, for $x >$ 0.5 a "Curie-Weiss" (CW) metallic state (localized Co spins with a CW-$\chi$ and a $T$-linear resistivity) is established \cite{Chou1,Huang1}. The large thermopower for $x \sim 0.6 - 0.8$ is due to enhanced electronic spin entropy, that gets suppressed upon applying a magnetic field \cite{Wang}. Interestingly, for $x \sim 0.66 - 0.75$, $\chi (T)$ fits a CW law with an effective Weiss temperature $\theta$ that increases from $\sim$ -50 K to -125 K \cite{Chou1}. This suggests increasing AF correlations with increasing $x$, for $x >$ 0.5, and a long-range ordered AF state is known for $x \sim$ 0.75, 0.82 \cite{Huang2, Bayrakci1}. Surprisingly, neutron inelastic scattering indicates ferromagnetic in-plane spin-fluctuations for Na$_{0.75}$CoO$_{2}$ \cite{Boothroyd}. A similar result is also known for NaNiO$_{2}$ \cite{Darie}. For Na$_{0.6-0.8}$CoO$_{2}$, the average local moment per Co site is estimated to be larger than that expected in a simple chemical picture of electron doping due to Na ions. The doped electrons  are expected to form low spin $S = 0$ Co$^{3+}$ sites in a background of $S = 1/2$ Co$^{4+}$ ions ($x$ = 0 implies all $S = 1/2$, Co$^{4+}$, while $x$ = 1 implies a lattice of $S = 0$, non-magnetic Co$^{3+}$ sites). In particular, the experimentally obtained average local moment for x = 0.75 is $\sim$1.1 $\mu$$_{B}$, while expected value is  $\sim$0.43 $\mu$$_{B}$ \cite{Chou1}.

Recent NMR measurements which probe the local spin indicate dynamic spin fluctuations and interlayer coupling, mediated by the Co-O-Na-O-Co exchange path \cite{Ning}. Another recent NMR study has concluded the existence of 3 types of Co ions in Na$_{0.7}$CoO$_{2}$ \cite{Mukhamedshin}. More intriguingly, two independent neutron scattering studies \cite{Bayrakci2, Helme} on high Na content  single crystals ($x = 0.75, 0.82$), corresponding to a very dilute lattice of S = 1/2 Co$^{4+}$ ions, show that obtained spin wave dispersions
can be accurately modelled only by assuming S = 1/2 Co$^{4+}$ moments at all Co sites. This suggested possible phase separation in high Na-content materials.  Both these studies obtain a ferromagnetic in-plane (J$_{\parallel}$) and an AF out of plane (J$_{\perp}$) exchange interaction of comparable magnitudes, indicative of 3-dimensional magnetic fluctuations. The studies  highlight several unconventional aspects: (i) J$_{\parallel}$ is comparable in magnitude to J$_{\perp}$, inspite of a layered 2D structure \cite{Bayrakci2,Helme} (ii) $\theta$ inferred from fitted exchange parameters in terms of a local-moment picture, namely -(3J$_{\parallel}$ + J$_{\perp}$), is positive, while $\theta$ obtained from $\chi$ is negative \cite{Bayrakci2}. The possibility of intermediate spins ($S = 1$) for Co $^{3+}$ was invoked to explain this behavior (iii) The observed sharp spin wave modes along the c-axis requires that, in a phase separated scenario, clusters should be aligned along c-axis over many layers \cite{Helme} (iv) A Wigner crystal model could be reconciled with observed spin-waves if Co$^{3+}$ ions have a moment similar in size to Co$^{4+}$ ions \cite{Helme}.

Another interesting aspect of A$_{x}$MO$_{2}$(A = Li, Na; M = Co, Ni) is the alkali metal ion (A)-ordering observed in this isostructural family. Electron diffraction (ED) studies on Na$_{x}$CoO$_{2}$ found ordered Na ion-Na vacancy superlattices, beyond the simple hexagonal structure \cite{Zandbergen}. It identified a novel structural principle: the presence of lines of Na-ions and vacancies (along the [110] direction), 
rather than a simple maximization of Na-Na ions in the layered structure. This behavior arises from a minimization of electrostatic energy, combined with a constraint of occupying Na ions on discrete lattice sites \cite{Zhang}. There exist 2 types of A sites in A$_{x}$MO$_{2}$: A1 (or Na1: above or below the M ions in the 2D triangular layer) and A2 (or Na2: above or below centroids of triangles formed by M ions). The Na1 site was shown to be 67 meV higher in energy than Na2, making Na2 the preferred site \cite{Zhang}. ED studies confirm that Na2 sites are preferentially occupied, while Na1 sites are dominantly empty, for all $x$ between 0.15-0.75. The only exception is the $x$ = 0.5 compound which has Na1 and Na2 sites equally occupied \cite{Huang1}. The calculated formation energies as a function of $x$ for Li$_{x}$NiO$_{2}$ \cite{Arroyo} and Na$_{x}$CoO$_{2}$ \cite{Zhang} show very similar behavior, identifying ion ordering at particular $x$ values.

Given these unconventional magnetic observations and the A-ion ordering in the A$_{x}$MO$_{2}$ series, we resorted to a simple model which aims to study the relation between doping fraction $x$, A-ion ordering, and temperature dependent magnetization, spin-entropy and heat capacity. We describe the M spins by an antiferromagnetic Ising model on a triangular lattice. The doped electrons are also described as Ising spins. While our model applies equally to the isostructural A$_{x}$MO$_{2}$ family, we mainly focus on Na$_{x}$CoO$_{2}$ for comparion with experiments.

We make the following ansatz: a doped electron occupies an A2 site and, by symmetry,  interacts equally with all three Co$^{4+}$ sites constituting a host triangle. It does not simply transform a single Co$^{4+}$ site to a Co$^{3+}$ site. This electron is assumed to couple antiferromagnetically with the three Co$^{4+}$ spins of the host triangle, and this makes the coupling between Co$^{4+}$ spins of the host triangle to become negligible. Thus, doping results in a local suppression of frustration (Figs. 1a and 1b). The ansatz may be viewed as dynamic fluctuations of spins due to doping, as concluded by NMR and neutron scattering measurements \cite{Ning, Mukhamedshin,Bayrakci2,Helme}. It yields a quantification of frustration and consistency with many experiments: neutron scattering, A-ion ordering, heat-capacity and also provides an explanation for enhanced spin entropy in the presence of low frustration.

\begin{figure}
\includegraphics[width={15cm}]{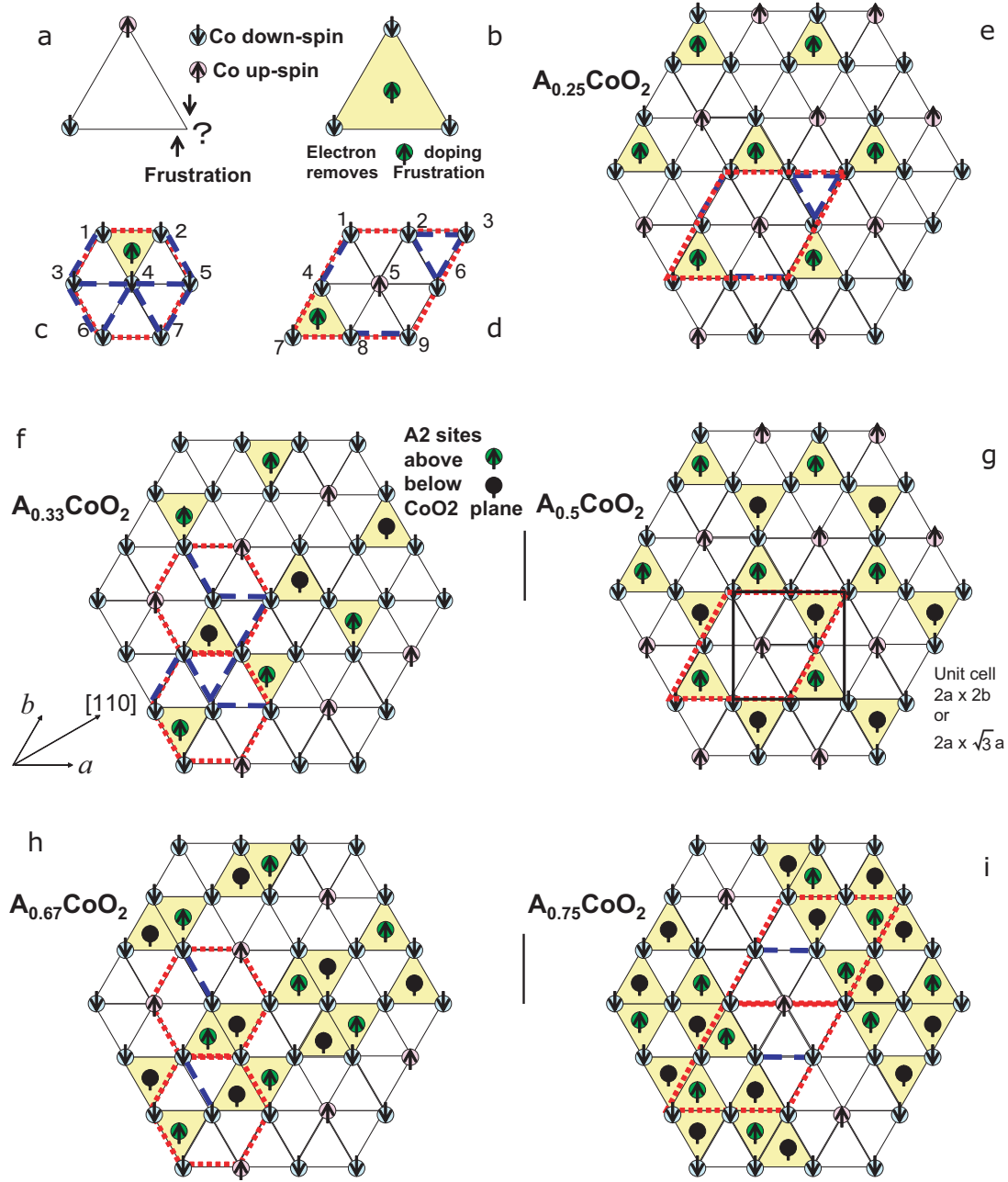}
\caption{(Color online) (a) Schematic of frustration, (b) ansatz: a doped electron
suppresses frustration by aligning three NN Co spins ferromagnetically, (c) hexagonal cell, (d) rhombic cell, (e)-(i)  Ordered structures obtained using the ansatz for $x~=~1/4,~1/3,~1/2,~2/3,~3/4$ (red dotted lines: lowest energy unit cells, blue dashed lines: unsatisfied bonds). For $x~=1/2$, black full line shows equivalent orthorhombic unit cell}
\label{fig:ssFig1.eps}
\end{figure} 

We investigate $x$-values which effectively represent the parent superconducting compositions (1/4, 1/3), the charge and spin-ordered compound (1/2), and the thermoelectrics (2/3, 3/4). These fractions can be classified on a geometric basis as (i) hexagonal fractions: x = 1/3 and 2/3 correspond to 1 and 2 occupied Na2 sites in a hexagonal cell, respectively (e.g. Fig. 1(c) for $x = 1/3$), and (ii) rhombic fractions: $x = 1/4, 1/2$ and $3/4$ correspond to occupying 1, 2 and 3 Na2 sites in the rhombic cell, respectively (e.g.Fig. 1(d) for $x = 1/4$). Lowest energy structures for these x-values obtained using the model are shown in Figs. 1(e) to 1(i). A doped electron from an Na2 site aligns the three Co spins of the host triangle antiparallel to it. The remaining Co spins
are also antiferromagnetically coupled Ising spins, aligned so as to minimize the total energy of the unit cells. The deduced Na-ion ordering for the lowest energy unit cells (red dotted lines in Figs. 1(e)-1(i))
match the experimentally obtained ion ordering for all x, except for x = 1/2. More significantly, it yields a quantification of frustration, defined as the number of unsatisfied bonds (blue dashed lines in Figs. 1(e) to 1(i)). The unsatisfied bonds are bonds aligned ferromagnetically inspite of the antiferromagnetic coupling. The bonds aligned ferromagnetically for host triangles (containing an A2 site) do not count as unsatisfied bonds. 
The fraction $f_{ub}(x)$ is the number of unsatisfied bonds normalised by the number of 
Co spins in a unit cell: $f_{ub}(x) = (3/4, 1, 0, 1/3, 1/4)$ for $x = (1/4, 1/3, 1/2, 2/3, 3/4)$ (Fig. 3(c)). Since $f_{ub}(0) = 1$, the fractions 1/4 and 1/3 are considered highly frustrated. 
The suppression of frustration is found to be a non-monotonic function of $x$. 
For $x = 1/3$ (Fig. 1(f)) and $2/3$ (Fig 1(h)), 
we need to use a two hexagon unit cell for the lowest energy structures; similarly for x = 3/4, the unit cell consists of two rhombii (Fig. 1(i)). 
However, for $x= 1/4$ (Fig. 1(e))  and $1/2$ (Fig. 1(g)), a single-rhombus unit cell is appropriate. 
These unit cells (Figs. 1(e)-1(i)) also correspond to the lowest $f_{ub}$ values, e.g. a single hexagon unit cell for
$x= 1/3$(Fig. 1c) leads to very high frustration with $f_{ub} = 2$. For $x=1$, the analysis shows all Co spins are ferromagnetically aligned with $f_{ub}(1)=0$; the unit cell in this case is a simple rhombus of two triangles that share a bond, with one doped electron per unit cell. It is noted that our model for x = 1 is related to the delta-star transformation of Baxter and Wu for the 2D triangular lattice, which was solved exactly in terms of three-spin interactions to yield a phase transition \cite{Baxter}.

\begin{figure}
\includegraphics[width={15cm}]{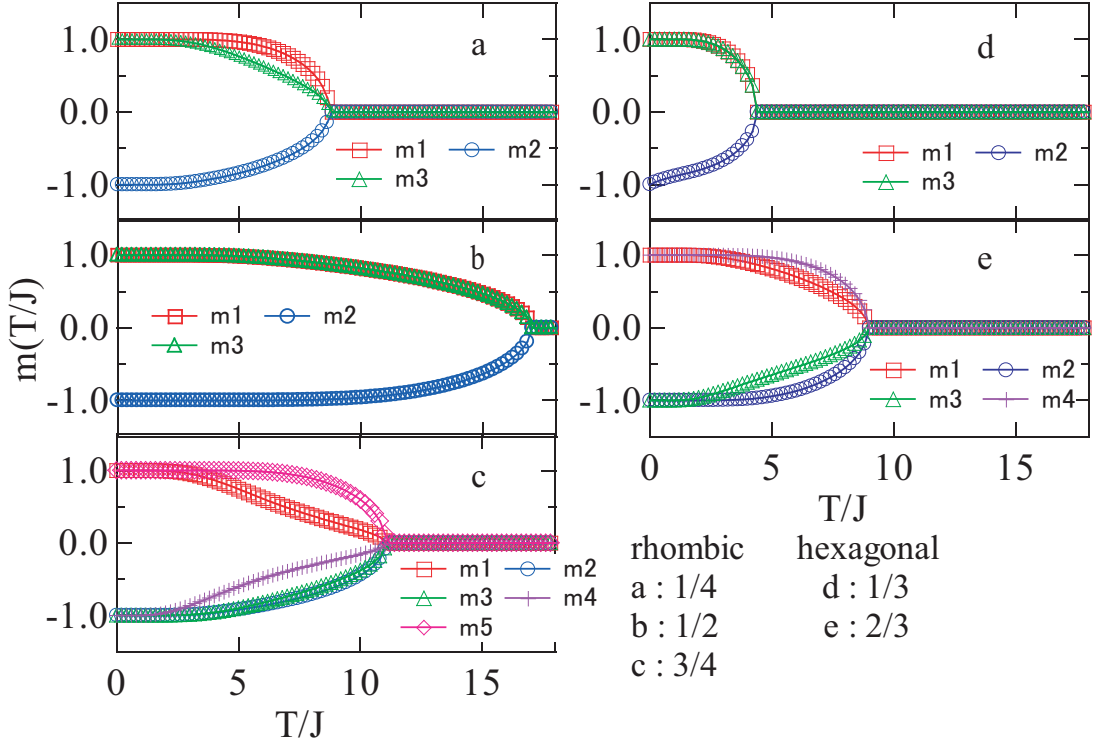}
\caption{(Color online) Magnetization $m_{i}(T)$, for $x~=~1/4,~1/3,~1/2,~2/3,~3/4$ ($J^{\prime}=1$), obtained from mean-field theory
for distinct sites $i$ in a unit cell.}
\label{fig:ssFig2}

\end{figure} 

The above discussion is at zero temperature, $T = 0$, with all fractions exhibiting complex magnetic long-range order. Since $f_{ub}$ varies non-monotonically with $x$, the magnetic transition temperatures, $T_c(x)$ are expected to show non-monotonic behaviour. It is also important to investigate $T$-dependent sub-lattice magnetization,
spin-entropy and heat capacity. We explore these aspects by performing a finite-temperature calculation using an Ising-like Hamiltonian, 

\begin{equation}
\label{eq:tempham1by4} 
H = \sum_{\langle ij \rangle}J_{ij}s_i s_j + J^{\prime}\sum_{k}\sum_{i_k}\sigma s_{i_k}.
\end{equation}

Here, $(ij)$ are nearest neighbour pairs of the unit cell. For proper counting, $J_{ij} = J/2$ whenever bond $(ij)$ is shared between two unit cells. Our ansatz implies that coupling between Co spins of a host triangle is zero, so that $J_{ij} = 0$ for all such pairs $(ij)$; and for all other pairs $(ij)$, $J_{ij} = J$. We choose energy units such that $J=1$. In the second term above, $k=1,2,...,n_{Na}$, $n_{Na}=xn_{Co}$ being the number of Na spins $\sigma$ ($n_{Co}$ is the number of Co spins $s$) in the unit cell; $i_k$ runs over the three vertices of host triangle $k$. The parameter $J^{\prime}$ is the Na-Co coupling. Eqn. 1 thus describes the unit cell Hamiltonian for any doping $x$. 

\begin{figure}
\includegraphics[width={15cm}]{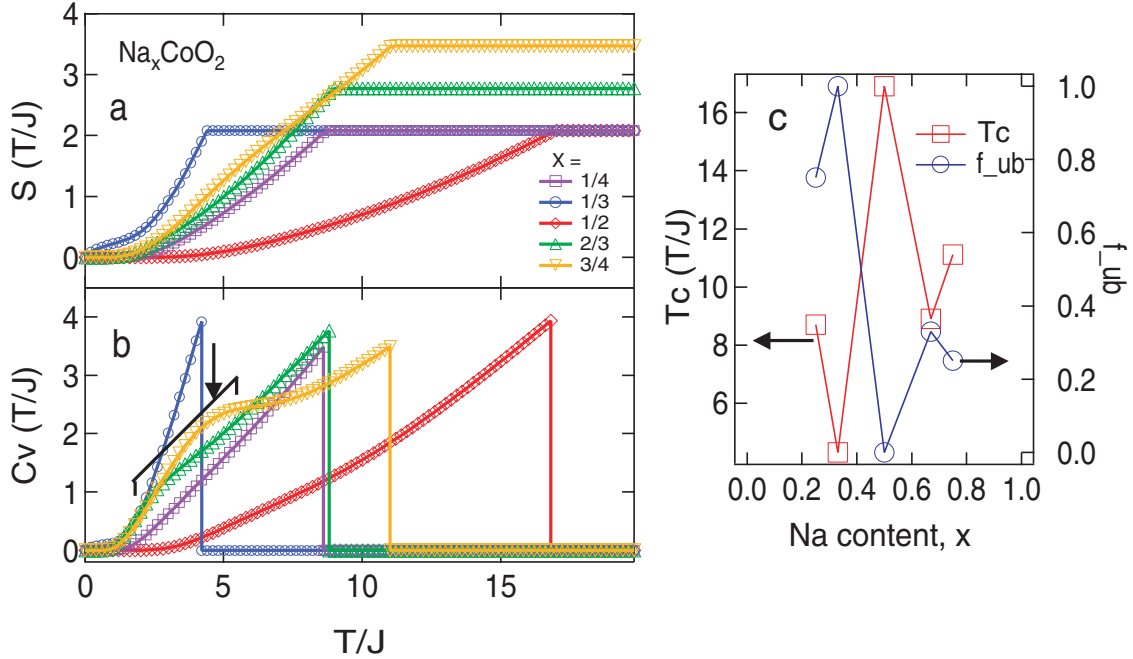}
\caption{(Color online) Plots of (a)  spin-entropy and (b) heat capacity as function of temperature(T/J), for $J^{\prime}=1$). (c) Plots of frustration $f_{ub}(x)$ and magnetic-ordering temperature $T_c(x)$ showing non-monotonic behaviour.  $f_{ub}(x)$ is maximum for $x=1/3$. $T_c(x)$ anticorrelates with $f_{ub}(x)$.}
\label{fig:ssFig3}
\end{figure}

We perform a mean-field calculation, making the approximation $s_i s_j \approx m_i s_j + m_j s_i - m_i m_j$, where magnetization $m_i = \langle s_i \rangle$. This calculation is straightforward, and we note that the doped electron spin, the Co spins neighbouring it, and those not neighbouring it, are distinct in mean-field theory. We solve self consistency equations for all the order parameters $m_i$. The number of distinct spins, $w(x)$, depends on the unit cell geometry and doping, and determines the high-temperature spin entropy $S(x, T>T_c(x)) = T \ln w(x)$. We calculate $T$-dependent $m_i$ at each distinct site, as well as spin-entropy $S$ and heat capacity $C_v$ for all fractions $x$. These thermodynamic quantities are presented in Figs. 2 and 3.

Fig. 2 shows magnetization $m(T)$ as a functon of $x$ for $J^{\prime}=1$ which indicates magnetic order occurs below
a critical temperature $T_c(x)$ for all $x$. The function $T_c(x)$ anticorrelates with $f_{ub}(x)$, and is also non-monotonic (Fig. 3(c)). The plots also suggest rhombic fractions are magnetically more stable than hexagonal fractions: in spite of larger frustration for $x=1/4$ than for $x=2/3$, the $T_c$ values are nevertheless comparable.

It is interesting to see that frustration $f_{ub}(x)$ is maximum for $x ~=~1/3$, which shows the highest superconducting $T_{sc}$ upon water intercalation \cite{Schaak}.
While we obtain magnetic order for $x ~=~ 1/3$, it has the lowest magnetic ordering $T_{c}$.
For $x = 1/4$, as well as $2/3$, the calculated magnetic $T_{c}'s$ are higher and comparable. 
While there is no report to date of observed magnetic ordering in
Na$_{x}$CoO$_{2}$ ($x = 1/4, 1/3$ and $2/3$), electrochemical measurements suggest an order-disorder transition 
around $x ~=~ 1/3$ \cite{Chou1}.  It is also known that Li$_{x}$CoO$_{2}$ and Na$_{x}$CoO$_{2}$
show ion-ordering at $x = 1/4, ~1/3, ~2/3$ or very close to these values \cite{Delmas,Zandbergen}. 
For $x = 1/3, 2/3$ and $3/4$, the Na-ion and vacancy ordering occurs along the [110] 
direction from our results (Fig 1(f), 1(h) and 1(i)). 
This is in good agreement 
with ED results for $x = 0.3, 0.64$ and $0.73$, which show incommensurate ordering \cite{Zandbergen}, while our results correspond to the nearest commensurate ordering. The results also match the ion ordering in $x = 1/4$ with a $(2a \times 2b)$ superlattice \cite{Arroyo,Zhang,Roger}. The obtained ion ordering 
for $x = 1/2$ gives a 'Star of David' or 'Kagome' structure within the triangular lattice \cite{Roger,Maekawa}.
This is not consistent with the $x = 1/2$ observed Na ion-order
\cite{Huang1}.  
Recent work shows that for $x = 1/2$, the formation of a divacancy cluster results in 
occupancy promotion of an Na2 site to an Na1 site \cite{Roger}. Since we have neglected Na1 sites in our model, we cannot address the Na1 site ion order. 
Nonetheless, the Co-lattice
magnetic structure obtained from our model for 
$x = 1/2$ ($2a \times 2b$ or equivalently, $2a \times \sqrt{3}a$ ; Fig. 1(g)) matches extended Hubbard model calculations \cite{Phillips} and recent neutron diffraction studies \cite{Yokoi,Gasparovic}, which show chains of antiferromagnetically ordered moments. While we obtain alternately anti-ferro and ferro chains parallel to each other,
ref. 21 models it as anti-ferro chains alternating with non-magnetic chains. However, they also point out that
the  S = 0 spin Co$^{3+}$ sites which are supposed to constitute non-magnetic chains may 
actually correspond to Co ions with fluctuating moments which do not order, consistent with
absence of unit-valence charge disproportionation \cite{Pickett,Gasparovic}.
The present results also show that the magnetic order is strongest for Na$_{0.5}$CoO$_{2}$, as is known experimentally, and frustration is completely suppressed ( $f_{ub}(x=1/2) = 0 $). For the larger doping fractions, $x~=~2/3, ~3/4$, the frustration is less than for $x ~=~ 1/3,~1/4$, but is still not zero. 

Fig.3 presents (a) spin-entropy($S$) and (b) heat capacity($C_{v}$) as functions of temperature(T/J). 
The plots show that  $C_{v}$ for $x=2/3$ and $3/4$ are anomalous (region marked by arrow in Fig. 3b). 
This anomalous behavior has been experimentally observed in the electronic part of 
$C_{v}$ for $x = 3/4$ \cite{Sales}. Also, Fig. 3a shows that 
changes in $S$ are higher for $x=2/3$ and $3/4$ compared with $x=1/4, ~1/3$ and $1/2$.
This is the doping regime which experimentally shows enhanced thermoelectricity that can by suppressed by an applied magnetic field, 
indicating a magnetic origin. Our results thus confirm the same. 
We investigated this anomaly for $x=3/4$ as a function of $J^{\prime}$, and found that this behaviour shows up only for $J^{\prime}>0.7$. If we associate $J^{\prime}$ and $J$ to be
$\approx$ J$_{\perp}$ and J$_{\parallel}$ obtained from neutron scattering \cite{Bayrakci2,Helme}, our results are consistent with the experimentally estimated $J_{\perp}$$~\simeq~ (0.73-2.1)J_{\parallel}$. 

In summary, we have presented a model for magnetism and related ion ordering for the
A$_{x}$MO$_{2}$ (A = Li, Na; M = Co,Ni) family of compounds, in which frustration and geometry play an important role. A mean-field calculation provides doping and temperature-dependent magnetization, spin-entropy
and heat capacity. Low doping ($x = 0.25, 0.33$) results in a highly frustrated regime.  A$_{0.5}$MO$_{2}$ has strongest order and no frustration, while high doping ($x = 0.66, 0.75$) leads to low frustration and higher spin-entropy in the model. The obtained results are in general agreement with neutron scattering, spin entropy-driven thermoelectricity and ion-ordering.

\end{document}